\documentclass[acus]{JAC2001}
\usepackage{graphicx}
\newcommand{\ud}{\mathrm{d}}
\begin{document}

\onecolumn 
\thispagestyle{empty}

\begin{center}
\begin{tabular}{p{130mm}}

\begin{center}
{\bf\Large
RMS/RATE DYNAMICS VIA LOCALIZED MODES} \\
\vspace{5mm}

{\bf\Large }

\vspace{1cm}

{\bf\Large Antonina N. Fedorova, Michael G. Zeitlin}\\

\vspace{1cm}

{\bf\large\it
IPME RAS, St.~Petersburg, 
V.O. Bolshoj pr., 61, 199178, Russia}\\
{\bf\large\it e-mail: zeitlin@math.ipme.ru}\\
{\bf\large\it http://www.ipme.ru/zeitlin.html}\\
{\bf\large\it http://www.ipme.nw.ru/zeitlin.html}
\end{center}

\vspace{1cm}

\abstract{ 
We consider some reduction from nonlinear 
Vlasov-Maxwell equation to rms/rate equations 
for second moments related quantities.  
Our analysis is based on va\-ri\-atio\-nal\--wa\-ve\-let approach to rational 
(in dynamical     
variables) approximation. It allows to           
control contribution from each scale of underlying multiscales                 
and represent solutions via multiscale exact nonlinear eigenmodes (waveletons)
expansions.        
Our approach provides the possibility to work with well-localized bases        
in phase space and best convergence properties of the corresponding                
expansions without perturbations or/and linearization procedures.  
}

\vspace{60mm}

\begin{center}
{\large Presented at the Eighth European Particle Accelerator Conference} \\
{\large EPAC'02} \\
{\large Paris, France,  June 3-7, 2002}
\end{center}
\end{tabular}
\end{center}
\newpage

\title{RMS/RATE DYNAMICS VIA LOCALIZED MODES}
\author{Antonina N.  Fedorova, Michael G. Zeitlin\\ 
IPME, RAS, V.O. Bolshoj pr., 61, 199178, St.~Petersburg, Russia
\thanks{e-mail: zeitlin@math.ipme.ru}\thanks{ http://www.ipme.ru/zeitlin.html;
http://www.ipme.nw.ru/zeitlin.html}   
}
\maketitle

\begin{abstract}
We consider some reduction from nonlinear 
Vlasov-Maxwell equation to rms/rate equations 
for second moments related quantities.  
Our analysis is based on va\-ri\-atio\-nal\--wa\-ve\-let approach to rational 
(in dynamical     
variables) approximation. It allows to           
control contribution from each scale of underlying multiscales                 
and represent solutions via multiscale exact nonlinear eigenmodes (waveletons)
expansions.        
Our approach provides the possibility to work with well-localized bases        
in phase space and best convergence properties of the corresponding                
expansions without perturbations or/and linearization procedures.                                   
\end{abstract}

\section{INTRODUCTION}
In this paper we consider the applications of a new nu\-me\-ri\-cal\--analytical 
technique based on the methods of local nonlinear harmonic
analysis or wavelet analysis to nonlinear rms/rate equations
for averaged quantities related to some particular case of nonlinear 
Vlasov-Maxwell equations.
Our starting point is a model and approach proposed by R. C. Davidson e.a. [1], [2].
According to [1] we consider electrostatic approximation for a thin beam.
This approximation is a particular important case of the general reduction
from statistical collective description based on Vlasov-Maxwell equations to a finite 
number of ordinary differential equations for the second moments related quantities
(beam radius and emittance). In our case these reduced rms/rate equations 
also contain some 
disribution averaged quantities besides the second moments, e.g. self-field energy
of the beam particles.
Such model is very efficient for analysis of many problems related to
periodic focusing accelerators, e.g. heavy ion fusion and tritium production.
So, we are interested in the understanding of collective properties, nonlinear dynamics and 
transport processes of intense non-neutral beams propagating through a periodic focusing 
field.
Our approach is based on the 
variational-wavelet approach from [3]-[16],[17]
that allows to consider rational type of 
nonlinearities in rms/rate dynamical equations containing  
statistically averaged quantities also.
The solution has the multiscale/multiresolution decomposition via 
nonlinear high-localized eigenmodes (waveletons),
which corresponds to the full multiresolution expansion in all underlying internal 
hidden scales. 
We may move
from coarse scales of resolution to the 
finest one to obtain more detailed information about our dynamical process.
In this way we give contribution to our full solution
from each scale of resolution or each time/space scale or from each nonlinear eigenmode. 
Starting from some electrostatic approximation of
Vlasov-Maxwell system and rms/rate dynamical 
models in part 2
we consider the approach based on
variational-wavelet formulation in part 3. 
We give explicit representation for all dynamical variables in the bases of
compactly supported wavelets or nonlinear eigenmodes.  Our solutions
are parametrized
by the solutions of a number of reduced standard algebraical problems.
We present also numerical modelling based on our analytical approach.

\section{RATE EQUATIONS}

In thin-beam approximation with negligibly small spread in axial 
momentum for beam particles we have in Larmor frame 
the following electrostatic approximation for Vlasov-Maxwell equations:
\begin{eqnarray}
&&\frac{\partial F}{\partial s}+x'\frac{\partial F}{\partial x}+
y'\frac{\partial F}{\partial y}-\Big(k(s)x+\frac{\partial\psi}{\partial x}\Big)
\frac{\partial F}{\partial x'}\\
&&-\Big(k(s)y+\frac{\partial\psi}{\partial y}\Big)
\frac{\partial F}{\partial y'}=0\nonumber
\end{eqnarray}
\begin{eqnarray}
&&\Big(\frac{\partial^2}{\partial x^2}+\frac{\partial^2}{\partial y^2}\Big)
\psi=-\frac{2\pi K}{N}\int\ud x'\ud y' F
\end{eqnarray}
where $\psi(x,y,s)$ is normalized electrostatic potential and
$F(x,y,x',y',s)$ is distribution function in transverse phase space 
$(x,y,x',y',s)$ with normalization
\begin{eqnarray}
N=\int\ud x\ud y n,\qquad n(x,y,s)=\int\ud x'\ud y' F
\end{eqnarray}
where $K$ is self-field perveance which measures self-field intensity [1].
Introducing self-field energy
\begin{eqnarray}
E(s)=\frac{1}{4\pi K}\int\ud x\ud y |\partial^2\psi/\partial x^2+
\partial^2\psi/\partial y^2 |
\end{eqnarray}
we have obvious equations for root-mean-square beam radius $R(s)$
\begin{eqnarray}
R(s)=<x^2+y^2>^{1/2}
\end{eqnarray}
and unnormalized beam emittance $\qquad\varepsilon^2(s)=$ 
\begin{equation}
4(<x'^2+y'^2><x^2+y^2>-<xx'-yy'>),
\end{equation}
which appear after averaging second-moments quantities 
regarding distribution function $F$:
\begin{equation}
\frac{\ud^2 R(s)}{\ud s^2}+\Big(k(s)R(s)-\frac{K(1+\Delta)}{2R^2(s)}\Big)R(s)=
\frac{\varepsilon^2(s)}{4R^3(s)}
\end{equation}
\begin{equation}
\frac{\ud\varepsilon^2(s)}{\ud s}+8R^2(s)\Big(\frac{\ud R}{\ud s}
\frac{K(1+\Delta)}{2R(s)}-\frac{\ud E(s)}{\ud s}\Big)=0,
\end{equation}
where the term $K(1+\Delta)/2$ may be fixed in some
interesting cases, but generally we have it only as average
\begin{equation}
K(1+\Delta)/2=-<x\partial\psi/\partial x+y\partial\psi/\partial y>
\end{equation}
regarding distribution $F$.
Anyway, the rate equations (7), (8) represent reasoanable reductions for the 
second-moments related quantities from the full nonlinear 
Vlasov-Poisson system. 
For trivial distributions Davidson e.a. [1] found additional reductions.
For KV distribution (step-function density) the second rate equation (8)
is trivial, $\varepsilon(s)$=const and we have only one nontrivial rate 
equation for rms beam radius (7).
The fixed-shape density profile ansatz for axisymmetric distributions
in [1] also leads to similar situation: emittance conservation and 
the same envelope equation with two shifted constants only.

\section{MULTISCALE REPRESENTATIONS}

Accordingly to our approach [3]-[16], [17] which allows us to find exact solutions 
as for Vlasov-like systems (1)-(3) as for rms-like systems (7),(8)
we need not to fix particular case of distribution function $F(x,y,x',y',s)$.
Our consideration is based on the following multiscale $N$-mode anzatz: 
\begin{eqnarray}
&&F^N(x,y,x',y',s)=\\
&&\sum^{N}_{i_1,\dots,i_5=1}a_{i_1,\dots,i_5}
\bigotimes^5_{k=1}A_{i_k}(x,y,x',y',s)\nonumber
\end{eqnarray}
\begin{equation}
\psi^N(x,y,s)=\sum^{N}_{j_1,j_2,j_3=1}b_{j_1,j_2,j_3}\bigotimes^3_{k=1}B_{j_k}(x,y,s)
\end{equation}
Formulae (10), (11) provide multiresolution representation
for variational solutions of system (1)-(3) [3]-[16],[17].
Each high-localized mode/harmonics  $A_j(s)$ corresponds
to level $j$ of resolution from the whole underlying infinite
scale of spaces:
$
\dots V_{-2}\subset V_{-1}\subset V_0\subset V_{1}\subset V_{2}\subset\dots,
$
where the closed subspace
$V_j (j\in {\bf Z})$ corresponds to  level j of resolution, or to scale j.
The construction of such tensor algebra based multiscales bases are considered in [17].
We'll consider rate equations (7), (8) as the following operator equation.
Let $L$, $P$, $Q$ be an arbitrary nonlinear (rational in dynamical variables) first-order 
matrix differential 
operators with matrix dimension d
(d=4 in our case) corresponding to the system of equations (7)-(8), 
which act on some set of functions
$\Psi\equiv\Psi(s)=\Big(\Psi^1(s),\dots,\Psi^d(s)\Big), \quad s \in\Omega\subset R$
from $L^2(\Omega)$: $Q(R,s) \Psi(s)=P(R,s)\Psi(s)$ or
\begin{equation}
L\Psi\equiv L(R,s)\Psi(s)=0
\end{equation}
where
$R\equiv R(s,\partial /\partial s, \Psi)$.
Let us consider now the N mode approximation for solution as the following 
expansion in some high-localized wavelet-like basis: 
\begin{equation}
\Psi^N(s)=\sum^N_{r=1}a^N_{r}\phi_r(s)
\end{equation}
We shall determine the coefficients of expansion from the following variational conditions
(different related variational approaches are considered in [3]-[16]):
\begin{equation}
L^N_{k}\equiv\int(L\Psi^N)\phi_k(s)\ud s=0
\end{equation}
We have exactly $dN$ algebraical equations for  $dN$ unknowns $a_{r}$.
So, variational approach reduced the initial problem (7), (8) to the problem of solution 
of functional equations at the first stage and some algebraical problems at the second
stage. 
As a result we have the following reduced algebraical system
of equations (RSAE) on the set of unknown coefficients $a_i^N$ of
expansion (14):
\begin{eqnarray}
H(Q_{ij},a_i^N,\alpha_I)=M(P_{ij},a_i^N,\beta_J),
\end{eqnarray}
where operators $H$ and $M$ are algebraization of RHS and LHS of initial problem
(12).
$Q_{ij}$ ($P_{ij}$) are the coefficients of LHS (RHS) of the initial
system of differential equations (7), (8) and as consequence are coefficients
of RSAE.
$I=(i_1,...,i_{q+2})$, $ J=(j_1,...,j_{p+1})$ are multiindexes, by which are
labelled $\alpha_I$ and $\beta_I$,  the other coefficients of RSAE (15):
\begin{equation}
\beta_J=\{\beta_{j_1...j_{p+1}}\}=\int\prod_{1\leq j_k\leq p+1}\phi_{j_k},
\end{equation}
where $p$ is the degree of polynomial operator $P$ (12)
\begin{equation}
\alpha_I=\{\alpha_{i_1}...\alpha_{i_{q+2}}\}=\sum_{i_1,...,i_{q+2}}\int
\phi_{i_1}...\dot{\phi_{i_s}}...\phi_{i_{q+2}},
\end{equation}
where $q$ is the degree of polynomial operator $Q$ (12),
$i_\ell=(1,...,q+2)$, $\dot{\phi_{i_s}}=\ud\phi_{i_s}/\ud s$.
According to [3]-[16] we may extend our approach to the case when we have additional
constraints as (3) on the set of our dynamical variables $\Psi$=\{$R$, $\varepsilon\}$
and additional averaged terms (4), (9) also.
In this case by using the method of Lagrangian multipliers we again may apply the same 
approach but
for the extended set of variables. As a result we receive the expanded system 
of algebraical equations
analogous to the system (15). Then, after reduction we again can extract from its 
solution the coefficients 
of expansion (13).  
It should be noted that if we consider only truncated expansion (13) with N terms
then we have the system of $N\times d$ algebraical equations
with the degree $\ell=max\{p,q\}$
and the degree of this algebraical system coincides
with the degree of the initial system.
So, after all we have the solution of the initial nonlinear
(rational) problem  in the form
\begin{eqnarray}
R^N(s)&=&R(0)+\sum_{k=1}^Na_k^N \phi_k(s)\\
\varepsilon^N(s)&=&\varepsilon(0)+\sum_{k=1}^Nb_k^N \phi_k(s)
\end{eqnarray}
where coefficients $a_k^N$, $b_k^N$ are the roots of the corresponding
reduced algebraical (polynomial) problem RSAE (15).
Consequently, we have a parametrization of the solution of the initial problem
by solution of reduced algebraical problem (15).
The problem of
computations of coefficients $\alpha_I$ (17), $\beta_J$
(16) of reduced algebraical
system
may be explicitly solved in wavelet approach.
The obtained solutions are given
in the form (18, (19),
where
$\phi_k(s)$ are proper wavelet bases functions (e.g., periodic or boundary). 
It should be noted that such representations 
give the best possible localization
properties in the corresponding (phase)space/time coordinates. 
In contrast with different approaches formulae (18), (19) do not use perturbation
technique or linearization procedures 
and represent dynamics via generalized nonlinear localized eigenmodes expansion.  
\begin{figure}[htb]                                                             
\centering                                                                       
\includegraphics*[width=60mm]{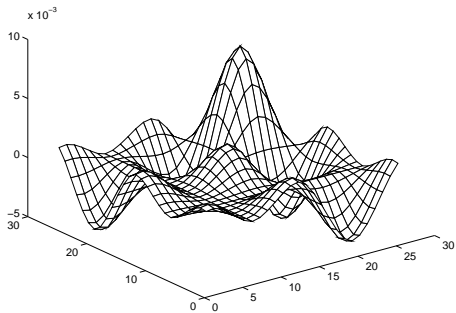}
\caption{Waveleton-like distribution.}                              
\end{figure}                                                                   
\begin{figure}[htb]                                                            
\centering                                                                      
\includegraphics*[width=60mm]{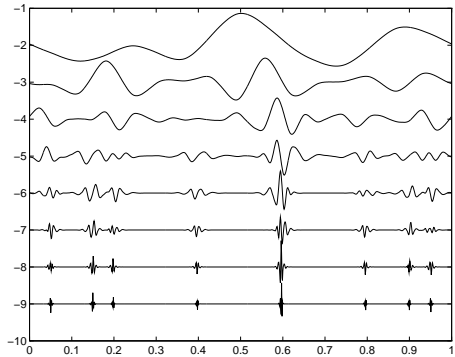}
\caption{Multiscale decomposition.}                                           
\end{figure}                                                                  
Our N mode construction (18), (19) gives the following  
general multiscale representation:
\begin{eqnarray*}
R(s)&=&R_{N}^{slow}(s)+\sum_{i\ge N}R^i(\omega_is),
\quad \omega_i \sim 2^i\\
\varepsilon(s)&=&\varepsilon_{N}^{slow}(s)+\sum_{j\ge N}\varepsilon^j(\omega_js),
\quad \omega_j \sim 2^j
\end{eqnarray*}
where $R^i(s), \varepsilon^j(s)$ are represented by some family of (nonlinear)
eigenmodes and gives the full multiresolution/multiscale representation in the
high-localized wavelet bases.
The corresponding decomposition is  presented on Fig.~2 and 
two-dimensional localized mode (waveleton) contribution to distribution function
is presented on Fig.1.
As a result we can construct different (stable) patterns from high-localized (coherent) 
structures in      
spa\-ti\-al\-ly\--ex\-te\-nd\-ed stochastic systems with complex collective behaviour.

 \end{document}